\documentclass[sigconf]{acmart}

\usepackage{multirow} 

\begin{document}
\title{Mitigating Shared-Private Branch Imbalance via Dual-Branch Rebalancing for Multimodal Sentiment Analysis}
\author{Chunlei Meng$^1$, Jiabin Luo$^2$, Pengbin Feng$^3$, Zhenglin Yan$^1$, Chengyin Hu$^4$, \\Zhongxue Gan$^1$, Chun Ouyang$^1$$*$}
\author{$^1$the College of Intelligent Robotics and Advanced Manufacturing, Fudan University\\$^2$Peking University, $^3$University of Southern California, $^4$China University of Petroleum-Beijing at Karamay}
\author{clmeng23@m.fudan.edu.cn, This study has been Accepted by ACM MM 2026.}




\begin{abstract} 
Multimodal Sentiment Analysis (MSA) requires integrating language, acoustic, and visual signals without sacrificing modality-specific sentiment evidence. Existing methods mainly improve either shared-private decomposition or cross-modal interaction. Although effective, both ultimately depend on how shared and modality-specific evidence is organized before prediction. We observe that, under standard shared-private pipelines, modality heterogeneity often induces a branch-imbalance process: dominant shared patterns accumulate in the shared branch, yielding redundant and modality-biased evidence, while repeated interaction and rigid alignment gradually leak shared information into modality-specific channels and weaken discriminative private representations. As a result, the complementarity between shared and private representations is reduced, limiting robust sentiment reasoning. To address this issue, we propose the Dual-Branch Rebalancing Framework (DBR) on top of a standard multimodal decoupling stage. In the shared branch, a Temporal-Structural Factorization (TSF) module disentangles temporal evolution from structural dependencies and adaptively integrates them to reduce shared redundancy. In the private branch, an Anchor-Guided Private Routing (AGPR) module preserves discriminative modality-specific patterns while allowing controlled cross-modal borrowing. A Bidirectional Rebalancing Fusion (BRF) module then reunifies the two regularized branches in a context-aware manner for final prediction. Extensive experiments on CMU-MOSI, CMU-MOSEI, and MIntRec demonstrate that DBR consistently outperforms the compared baselines. Further analyses show that these improvements come from coordinated mitigation of branch imbalance. 
\end{abstract} 

\ccsdesc[500]{Computing methodologies~Machine learning} \ccsdesc[300]{Computing methodologies~Natural language processing} \ccsdesc[300]{Information systems~Multimedia information systems} \keywords{multimodal sentiment analysis, multimodal representation learning, multimodal fusion, shared-private decomposition, affective computing} 
\maketitle 
\begin{figure}[htbp] 
\centering 
\includegraphics[width=1\linewidth]{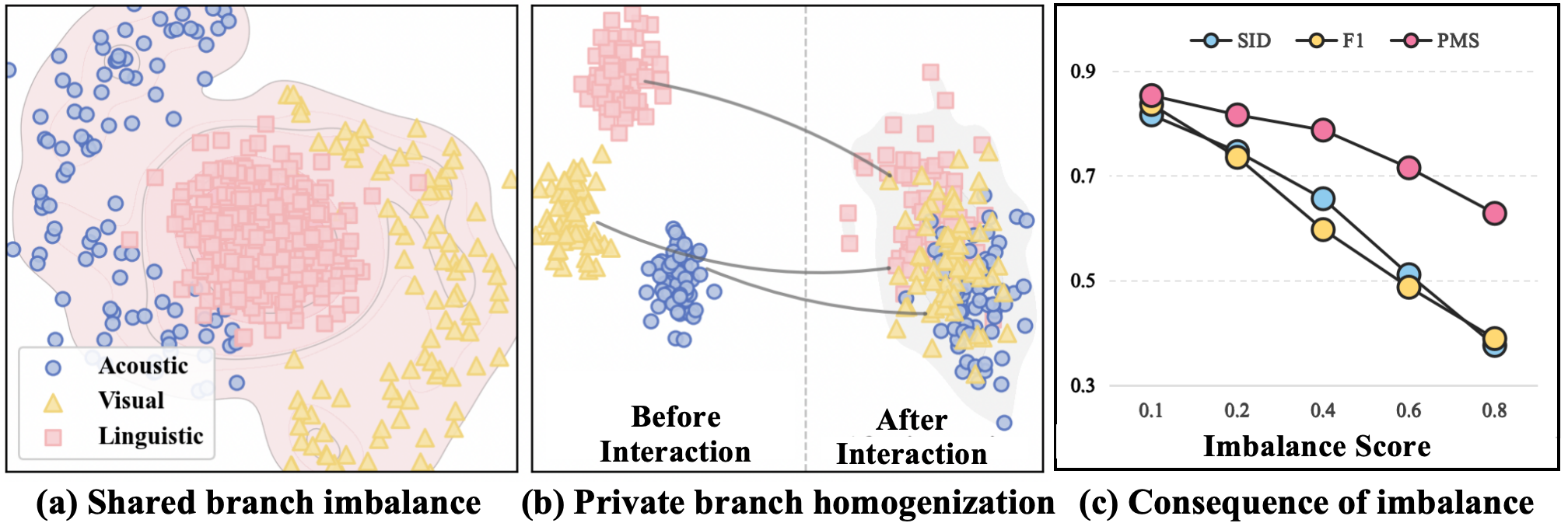} 
\caption{Diagnosis of branch imbalance in a standard shared-private multimodal pipeline. (a) Shared branch. Cross-modal patterns collapse into a modality-dominated region, causing shared redundancy. (b) Private branch. Initially separable representations severely overlap after cross-modal interaction, leading to progressive homogenization. (c) As the sample-level imbalance increases, the overall F1, Shared Information Diversity (SID), and Private Modality Separability (PMS) all consistently decrease. These findings demonstrate that branch imbalance aggravates shared redundancy, dilutes private specificity, and harms downstream performance.}
\label{fig:motivation} 
\end{figure}

\section{Introduction} 
Multimodal Sentiment Analysis (MSA) aims to infer human sentiment and emotion by jointly reasoning over linguistic, acoustic, and visual signals~\cite{TFN,misa,GCL}. As a representative task in affective computing and human-centered multimodal intelligence, MSA has achieved substantial progress in recent years. Yet its central obstacle remains the inherent heterogeneity across modalities: the three modalities differ in signal form, temporal dynamics, semantic granularity, and noise sensitivity~\cite{d2r,MGJR,DEVA}. 

To cope with modality heterogeneity, existing methods mainly advance along two routes, but both revolve around the same core tension: how to organize shared evidence versus modality-specific evidence before prediction. The first route makes this split explicit through shared-private decomposition. MISA~\cite{misa} projects each modality into invariant and specific subspaces to reduce modality gaps while preserving characteristic cues. FDMER~\cite{FDMER} learns shared and characteristic representations. ConFEDE~\cite{confede} decomposes each modality into similarity and dissimilarity features through contrastive feature decomposition, while DMD~\cite{dmd} performs graph distillation separately on decoupled modality-irrelevant and modality-exclusive spaces. More recent models refine the granularity of decoupling itself: TSDA~\cite{TSDA} separates temporal dynamics from spatial/structural context before interaction, CLCR~\cite{CLCR} organizes cross-modal learning in a multi-level shared-private hierarchy and explicitly restricts exchange to shared channels, and TSD~\cite{TSD} introduces pairwise shared and private representations to model multi-granular sentiment cues. The second route mainly strengthens cross-modal interaction around these representations, using cross-modal attention, routing, experts, or language-centered aggregation as in MAG-BERT~\cite{MAG-BERT}, D2R~\cite{d2r}, DEVA~\cite{DEVA}, EMOE~\cite{EMOE}, and DLF~\cite{DLF}. More related works see supplementary material. These advances substantially improve MSA, but they are still mostly discussed from the perspective of cleaner disentanglement or stronger fusion. Recent studies have already exposed fragments of a deeper difficulty. DLF~\cite{DLF} points out that directly extracting shared information or naively fusing modalities can introduce redundancy and conflicts because modalities are treated equally and information is mutually transferred across modality pairs. EMOE~\cite{EMOE} highlights two coupled symptoms (modality balance dilemma and modality specialization disappearance) showing that dominant modalities can steer learning while modality-specific predictive ability is weakened. 

Although these observations are insightful, they are still discussed separately and do not provide a unified explanation of how modality heterogeneity continues to influence representation learning after shared-private decomposition. To further investigate this phenomenon and its impact, we conduct a lightweight motivating diagnosis using a standard shared-private multimodal sentiment analysis pipeline. Fig.~\ref{fig:motivation}(a) shows that the shared branch does not organize common evidence in a balanced manner. Instead, recurring cross-modal patterns collapse into a dense common region that is visibly dominated by one modality, such as high-frequency lexical markers, repeated visual frames, or stable prosodic structure. This suggests that the shared branch tends to over-absorb dominant common patterns, resulting in redundancy-heavy and modality-biased shared representations. Fig.~\ref{fig:motivation}(b) further shows that, although private representations are relatively separable before interaction, conventional cross-modal interaction progressively draws them toward a more overlapping and homogeneous region. This indicates that modality-specific structure is gradually diluted, as private channels drift toward leaked common semantics rather than preserving distinctive evidence.

These observations suggest that the seemingly isolated symptoms can be understood in a unified way as a heterogeneity-induced imbalance between the shared and private branches. That is, shared-private decomposition does not remove modality heterogeneity, but redistributes it unevenly across the two branches. We refer to this phenomenon as the Shared-Private Branch Imbalance problem. To further reveal its consequence, we sort test samples by a sample-level branch imbalance score and divide them into several bins from low to high imbalance. As shown in Fig.~\ref{fig:motivation}(c), as the imbalance level increases, the overall F1, the Shared Information Diversity (SID) measured by the normalized effective rank of shared representations, and the Private Modality Separability (PMS) measured by the silhouette score of private representations all decrease consistently. This result shows that branch imbalance is not merely a representational artifact. Instead, it progressively aggravates shared redundancy, weakens private specificity, reduces the complementarity between shared and private evidence, and ultimately degrades downstream prediction quality. (Detailed information see supplementary material). This diagnosis also clarifies the specific form of the problem. In the shared branch, dominant recurring common patterns are easily over-accumulated, making the shared representation redundancy-heavy, biased toward dominant modalities, and less sensitive to subtle affective cues. In the private branch, repeated cross-modal borrowing, shallow aggregation, or rigid alignment can progressively weaken modality-specific structure, causing private channels to drift toward common semantics and become increasingly homogeneous. The net effect is reduced complementarity between shared and private evidence, limited preservation of modality-specific expressiveness, and weakened robustness in downstream sentiment reasoning.

Based on this view, we propose Dual-Branch Rebalancing Framework (DBR), a dual-branch multimodal framework built on a standard multimodal decoupling stage. Specifically, the shared branch employs Temporal-Structural Factorization (TSF) to separate temporal and structural evidence and then adaptively rebalance them to suppress shared redundancy, while the private branch uses Anchor-Guided Private Routing (AGPR) to preserve discriminative modality-specific cues and regulate controlled cross-modal borrowing. The two branches are finally unified by a Bidirectional Rebalancing Fusion (BRF) module, which adaptively integrates shared and private evidence for prediction. In this way, DBR is not simply another stronger fusion block or a finer-grained decomposition model; rather, it is a coordinated branch-rebalancing framework targeted at the two coupled manifestations of the shared-private imbalance process. The main contributions are as follows: 
\begin{enumerate}
    \item We identify the Shared-Private Branch Imbalance problem in MSA, showing that modality heterogeneity is not eliminated by shared-private decomposition but redistributed unevenly across the two branches. To address this issue, we propose the Dual-Branch Rebalancing Framework (DBR) built on a standard multimodal decoupling stage.
    
    \item We design two branch-specific modules to tackle the coupled manifestations of this problem. Temporal-Structural Factorization (TSF) suppresses redundancy accumulation in the shared branch by disentangling and rebalancing temporal and structural evidence, while Anchor-Guided Private Routing (AGPR) preserves discriminative modality-specific cues under controlled cross-modal borrowing to mitigate private-feature dilution.
    
    \item We further propose Bidirectional Rebalancing Fusion (BRF) module to adaptively integrate the two rebalanced branches for representation learning and final prediction. 
\end{enumerate}

\begin{figure*}[htbp] 
\centering 
\includegraphics[width=1\linewidth]{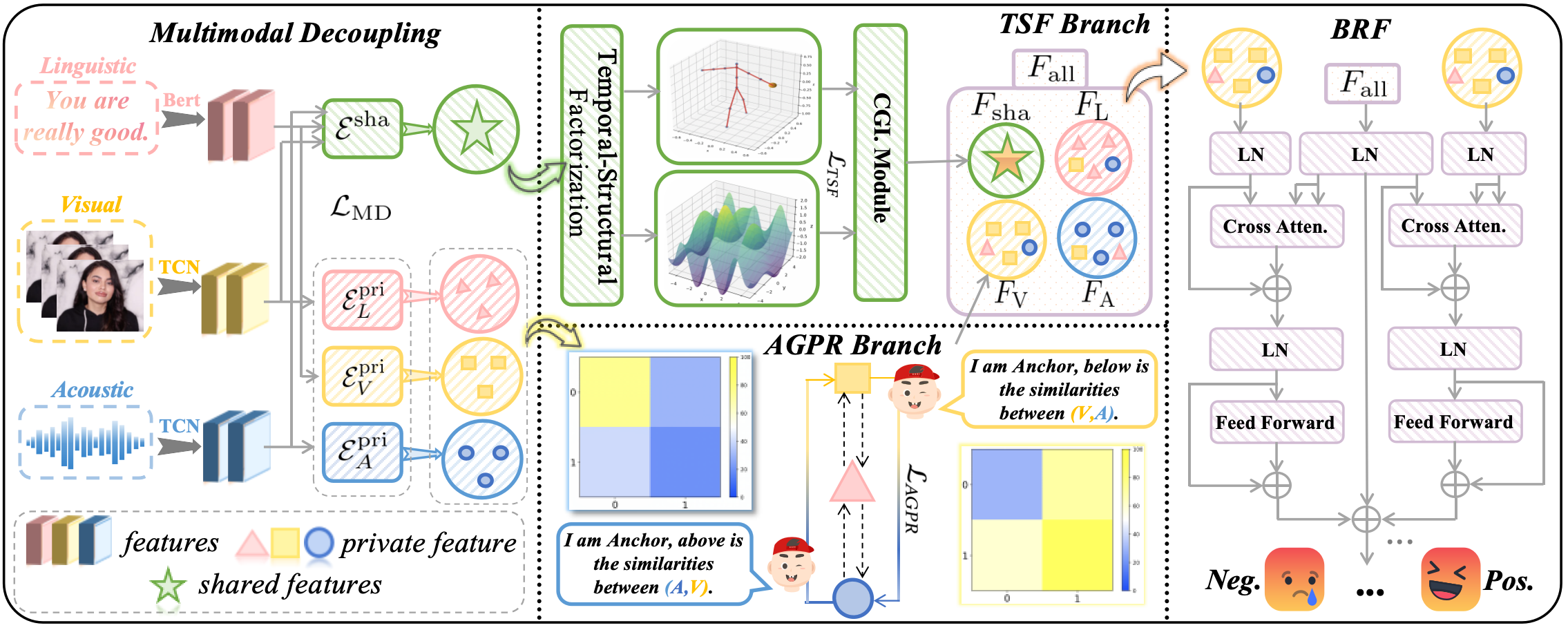} \caption{Architecture illustration of DBR. The left shows the multimodal decoupling . The center  highlights the TSF branch and the AGPR branch. Only interactions between the (yellow) visual and (blue) acoustic modalities are shown, with analogous cases for other modalities indicated by dotted lines. The right presents the BRF, exemplifying the fusion of the (yellow) visual modality with $F_{\text{all}}$. Similar fusion processes for other modalities are omitted.} \Description{An overview diagram of the full DBR pipeline. The diagram shows the initial multimodal decoupling stage, a shared branch with Temporal-Structural Factorization, a private branch with anchor-guided private routing, and a final bidirectional fusion module for prediction.} 
\label{fig:MESC} 
\end{figure*}

\section{Proposed Method}
\label{sec:SDAG}
As shown in Fig.~\ref{fig:MESC}, DBR is built on a standard Multimodal Decoupling (MD) stage and then organized as a dual-branch pipeline for mitigating \emph{heterogeneity-induced shared-private branch imbalance}. After shared-private initialization, this imbalance has two coupled manifestations: over-concentration of dominant temporal/structural patterns in the shared branch, and progressive dilution of modality-specific cues in the private branch. TSF addresses the former, AGPR addresses the latter, and BRF reunifies the two branches in a context-aware manner for prediction. The design goal is not to maximize architectural complexity, but to expose the imbalance process, regulate its two manifestations with dedicated modules, and then fuse the two branches only after they have been separately regularized.

\subsection{Multimodal Decoupling}

As a standard starting point, multimodal decoupling separates information that should be shared across modalities from information that should remain modality-specific. Without this first split, later redundancy reduction and private-feature preservation would operate on already entangled features and become harder to interpret.

Given input streams from visual (V), acoustic (A), and linguistic (L) modalities, we first encode visual and acoustic features using independent Temporal Convolutional Networks (TCNs)~\cite{TCN}, and extract linguistic features via a pre-trained BERT encoder. Let $\mathbf{X}_m \in \mathbb{R}^{T_m \times d}$ represent the feature sequence for modality $m \in \{V, A, L\}$, where $T_m$ is the sequence length and $d$ is the feature dimension.

To disentangle modality-invariant and modality-specific factors, we employ a decoupling encoder for each modality, consisting of a shared encoder $\mathcal{E}_m^{\text{sha}}$ and a private encoder $\mathcal{E}_m^{\text{pri}}$:
\begin{equation}
\mathbf{X}_m^{\text{sha}} = \mathcal{E}_m^{\text{sha}}(\mathbf{X}_m), \quad
\mathbf{X}_m^{\text{pri}} = \mathcal{E}_m^{\text{pri}}(\mathbf{X}_m).
\label{eq:md-enc}
\end{equation}

\textbf{Decorrelation orthogonality loss.}
To encourage complementary information between shared and private representations, we introduce a decorrelation loss. For each modality $m$, we compute the empirical cross-covariance between zero-mean features $\tilde{\mathbf{X}}_m^{\text{sha}}, \tilde{\mathbf{X}}_m^{\text{pri}} \in \mathbb{R}^{T_m \times d}$ (centered along the temporal dimension):
\begin{equation}
\mathrm{Cov}(\mathbf{X}_m^{\text{sha}}, \mathbf{X}_m^{\text{pri}})
= \frac{1}{T_m - 1} \big(\tilde{\mathbf{X}}_m^{\text{sha}}\big)^\top
\tilde{\mathbf{X}}_m^{\text{pri}} \in \mathbb{R}^{d \times d}.
\end{equation}
We then penalize the off-diagonal elements:
\begin{equation}
\mathcal{L}_{\mathrm{MD}} =
\sum_{m \in \{V, A, L\}}
\left\| \mathrm{Off} \big( \mathrm{Cov}(\mathbf{X}_m^{\text{sha}}, \mathbf{X}_m^{\text{pri}}) \big) \right\|_1,
\label{eq:md-de-corr}
\end{equation}
where $\mathrm{Off}(\cdot)$ zeros the diagonal and $\|\cdot\|_1$ is the entry-wise $\ell_1$ norm. This encourages near-orthogonality between shared and private features, reducing redundancy and improving representation diversity.

\subsection{Temporal-Structural Factorization Branch}

Not all shared information contributes equally to sentiment reasoning. Temporal evolution captures how sentiment unfolds over time, while structural dependence in our setting refers to intra-sequence contextual organization rather than pixel-level geometry. Under branch imbalance, these factors can be mixed into one shared representation, allowing repeated frames or dominant tokens to over-occupy the shared branch and accumulate redundant evidence. Inspired by the Information Bottleneck (IB) principle~\cite{IB}, we therefore seek shared representations that remain predictive of labels while suppressing unnecessary cross-modal redundancy.

Let $\mathbf{X} = \{\mathbf{X}_m\}_{m=1}^M$ be multimodal features, $Y$ the emotion label, and $\mathbf{Z}$ the joint representation. We optimize
\begin{equation}
\max_{p(\mathbf{z}|\mathbf{x})} I(\mathbf{Z}; Y) - \beta I(\mathbf{Z}; \mathbf{X}),
\end{equation}
where $\beta$ is a positive trade-off parameter. Motivated by this view, we factorize the shared branch into temporal and structural components and then rebalance them through a cross-stream gating module.

\textbf{Temporal-Structural Factorization.}
For each modality input $\mathbf{X}_m^{\text{sha}}$, we employ a Bidirectional LSTM (BiLSTM) to capture temporal dependencies in both directions:
\begin{align}
\mathbf{H}_m^{\mathrm{temp}} &= \mathrm{LN}\big(\mathrm{Linear}(\mathrm{BiLSTM}(\mathbf{X}_m^{\mathrm{sha}}))\big),
\end{align}
where $\mathbf{H}_m^{\mathrm{temp}} \in \mathbb{R}^{T_m \times d}$ denotes the temporally encoded representation.

Structural dependencies are modeled by multi-head self-attention over each sequence. We set
\begin{equation}
\mathbf{Q}_m = \mathbf{K}_m = \mathbf{V}_m = \mathbf{X}_m^{\mathrm{sha}},
\end{equation}
and compute
\begin{align}
\mathbf{H}_m^{\mathrm{struct}} =
\mathrm{LN}\big(\mathrm{Linear}(\mathrm{MultiHead}(\mathbf{Q}_m,\mathbf{K}_m,\mathbf{V}_m))\big),
\end{align}
where $\mathbf{H}_m^{\mathrm{struct}} \in \mathbb{R}^{T_m \times d}$ encodes structural information within each sequence.

\textbf{Cross-Stream Gating Integration (CGI).}
Once all temporal ($\mathbf{H}_m^{\mathrm{temp}}$) and structural ($\mathbf{H}_m^{\mathrm{struct}}$) features are obtained, we concatenate them across modalities:
\begin{equation}
\mathbf{Z}^{\mathrm{temp}} = \mathrm{Concat}\big(\mathbf{H}_m^{\mathrm{temp}}\big)_{m=1}^M, \quad
\mathbf{Z}^{\mathrm{struct}} = \mathrm{Concat}\big(\mathbf{H}_m^{\mathrm{struct}}\big)_{m=1}^M.
\end{equation}
CGI generates $K$ candidate fusion representations via branch-specific MLPs:
\begin{equation}
    \mathbf{f}_k = \mathrm{MLP}_k\big([\mathbf{Z}^{\mathrm{temp}}; \mathbf{Z}^{\mathrm{struct}}]\big),
\end{equation}
where $[\cdot;\cdot]$ denotes concatenation. Gating weights are computed as
\begin{equation}
    [g_1, \ldots, g_K] = \mathrm{Softmax}\big(\mathrm{MLP}_g([\mathbf{Z}^{\mathrm{temp}}; \mathbf{Z}^{\mathrm{struct}}])\big),
\end{equation}
where $\mathrm{MLP}_g$ is a lightweight network and $\sum_{k=1}^K g_k = 1$. The fused representation in the shared temporal-structural branch is
\begin{equation}
    \mathbf{Z}^{\mathrm{fusion}} = \sum_{k=1}^{K} g_k \cdot \mathbf{f}_k.
\end{equation}

\textbf{Temporal-Structural Decorrelation Loss.}
We further regularize temporal and structural features with two terms.

The temporal-structural decorrelation loss is
\begin{equation}
\mathcal{L}_{\mathrm{decor}} = \frac{1}{M}\sum_{m=1}^{M}
\big\|\mathrm{Corr}(\mathbf{H}_m^{\mathrm{temp}}, \mathbf{H}_m^{\mathrm{struct}})\big\|_F^2,
\end{equation}
where $\|\cdot\|_F$ denotes the Frobenius norm and $\mathrm{Corr}(\cdot,\cdot)$ is the cross-correlation matrix computed along the temporal dimension. Given centered features $\tilde{\mathbf{A}}, \tilde{\mathbf{B}} \in \mathbb{R}^{T_m \times d}$, the $(i,j)$-th entry is
\begin{equation}
\mathrm{Corr}(\mathbf{A},\mathbf{B})_{ij} =
\frac{\sum_{t=1}^{T_m} \tilde{A}_{t,i} \tilde{B}_{t,j}}
{\sqrt{\sum_{t=1}^{T_m} \tilde{A}_{t,i}^2}\,
 \sqrt{\sum_{t=1}^{T_m} \tilde{B}_{t,j}^2}}.
\end{equation}
Minimizing $\mathcal{L}_{\mathrm{decor}}$ encourages temporal and structural features to be weakly correlated.

The alignment loss constrains modality-wise means to be close to modality-agnostic global means:
\begin{equation}
\mathcal{L}_{\mathrm{align}} =
\frac{1}{M} \sum_{m=1}^{M}
\left( \big\|\boldsymbol{\mu}_m^{\mathrm{temp}} - \bar{\boldsymbol{\mu}}^{\mathrm{temp}}\big\|_2^2 +
       \big\|\boldsymbol{\mu}_m^{\mathrm{struct}} - \bar{\boldsymbol{\mu}}^{\mathrm{struct}}\big\|_2^2 \right),
\end{equation}
where $\boldsymbol{\mu}_m^{\mathrm{temp}}$ and $\boldsymbol{\mu}_m^{\mathrm{struct}}$ are the mean feature vectors of $\mathbf{H}_m^{\mathrm{temp}}$ and $\mathbf{H}_m^{\mathrm{struct}}$ over all time steps and samples in the mini-batch, and $\bar{\boldsymbol{\mu}}^{\mathrm{temp}}$ and $\bar{\boldsymbol{\mu}}^{\mathrm{struct}}$ are their averages over modalities. The TSF loss is
\begin{equation}
\mathcal{L}_{\mathrm{TSF}} = \alpha_1 \mathcal{L}_{\mathrm{decor}} + \alpha_2 \mathcal{L}_{\mathrm{align}},
\end{equation}
with $\alpha_1$ and $\alpha_2$ as balancing coefficients.

\subsection{Anchor-Guided Private Routing Branch}

The private branch should preserve what is unique to each modality, but complete isolation is also undesirable because complementary cues can still be useful. Under the same branch imbalance, however, naive cross-modal borrowing can gradually dilute discriminative private patterns and make the private branch more homogeneous. AGPR addresses this trade-off by using modality anchors to regulate who borrows information from whom and by how much.

Let $\mathcal{M} = \{L, A, V\}$ denote the set of modalities. For each modality $m \in \mathcal{M}$, we obtain a pooled private representation from $\mathbf{X}_m^{\mathrm{pri}}$ through a private encoder $f_m^{\mathrm{pri}}$:
\begin{equation}
\mathbf{z}_m^{\mathrm{pri}} = f_m^{\mathrm{pri}}(\mathbf{X}_m^{\mathrm{pri}}), \quad
\mathbf{z}_m^{\mathrm{pri}} \in \mathbb{R}^d.
\end{equation}
Each modality is assigned a trainable anchor $\mathbf{b}_m \in \mathbb{R}^d$, serving as a prototype in the private representation space. The similarity between modality $n$ and the anchor of modality $m$ is measured by cosine similarity:
\begin{equation}
s_{n \to m} =
\frac{\langle \mathbf{z}_n^{\mathrm{pri}}, \mathbf{b}_m \rangle}
{\|\mathbf{z}_n^{\mathrm{pri}}\|_2 \, \|\mathbf{b}_m\|_2}, \quad n \neq m.
\end{equation}
Larger values indicate that the private representation of modality $n$ contains information that is useful to modality $m$.

We compute incoming routing weights with a temperature-controlled softmax:
\begin{equation}
w_{n \to m} = \frac{\exp(\gamma s_{n \to m})}
{\sum_{k \neq m} \exp(\gamma s_{k \to m})},
\end{equation}
where $\gamma$ is a temperature hyperparameter. The updated private feature of modality $m$ is
\begin{equation}
\hat{\mathbf{z}}_m^{\mathrm{pri}} =
\mathbf{z}_m^{\mathrm{pri}} +
\lambda \sum_{n \neq m} w_{n \to m} \, \mathbf{z}_n^{\mathrm{pri}},
\end{equation}
where $\lambda$ is a feature sharing coefficient. This enables each modality to selectively absorb complementary private information while maintaining its own discriminative identity.

To ensure discriminability and independence of private features, we employ an anchor-based regularization loss with two parts. The alignment term pulls each private representation toward its anchor:
\begin{equation}
\mathcal{L}_{\mathrm{ali}} =
\sum_{m \in \mathcal{M}} \big\| \mathbf{z}_m^{\mathrm{pri}} - \mathbf{b}_m \big\|_2^2,
\end{equation}
and the separation term keeps private representations of different modalities apart:
\begin{equation}
\mathcal{L}_{\mathrm{sep}} =
\sum_{m \in \mathcal{M}} \sum_{n \neq m}
\max\left(
0,\,
\delta + \big\|\mathbf{z}_m^{\mathrm{pri}} - \mathbf{b}_m\big\|_2^2
        - \big\|\mathbf{z}_m^{\mathrm{pri}} - \mathbf{b}_n\big\|_2^2
\right),
\end{equation}
where $\delta$ is a margin hyperparameter. The AGPR regularization loss is
\begin{equation}
\mathcal{L}_{\mathrm{AGPR}} =
\beta_1 \mathcal{L}_{\mathrm{ali}} +
\beta_2 \mathcal{L}_{\mathrm{sep}},
\end{equation}
with $\beta_1$ and $\beta_2$ as trade-off weights.

\begin{table*}[t]
\caption{Performance Comparison on MOSI and MOSEI. $\uparrow$ and $\downarrow$ indicate that higher or lower value is better. Bold means best.}
\centering
\setlength{\tabcolsep}{2.6pt} 
\renewcommand{\arraystretch}{1.0} 
\begin{tabular}{l|ccccc|ccccc}
\toprule
\multirow{2}{*}{\textbf{Models}} 
& \multicolumn{5}{c|}{\textbf{CMU-MOSI}} 
& \multicolumn{5}{c}{\textbf{CMU-MOSEI}} \\
 & MAE ($\downarrow$) & Corr ($\uparrow$) & Acc-2(\%) & Acc-7(\%)& F1(\%)
 & MAE ($\downarrow$) & Corr ($\uparrow$) & Acc-2(\%) & Acc-7(\%) & F1(\%) \\
\midrule
\midrule

TFN~\cite{TFN} & 0.947 & 0.673 & 77.99 & 31.9 &77.95&0.572&0.714&78.5&51.6&78.96\\
LMF~\cite{LMF} & 0.950 & 0.651 & 77.90& 33.82&77.80&0.575&0.714&80.54&51.59&80.94\\
MuLT~\cite{MuLT}& 0.846 & 0.725 & 81.70 & 40.05 & 81.66 & 0.673 & 0.677 & 80.85 & 48.37 & 80.86 \\
PMR~\cite{PMR}& 0.895 & 0.689 & 79.88 & 40.60 & 79.83 & 0.645 & 0.689 & 81.57 & 48.88 & 81.56 \\
MISA~\cite{misa}     & 0.788 & 0.744 & 82.07 & 41.27 & 82.43 & 0.594 & 0.724 & 82.03 & 51.43 & 82.13 \\
Self-MM~\cite{self-mm}& 0.765 & 0.764 & 82.88 & 42.03 & 83.04 & 0.576 & 0.732 & 82.43 & 52.68 & 82.47 \\
FDMER~\cite{FDMER}& 0.760 & 0.777 & 83.01 & 42.88 & 83.22 & 0.571 & 0.743 & 83.88 & 53.21 & 83.35 \\
DMD~\cite{dmd}& 0.744 & 0.788 & 83.24 & 43.88 & 83.55 & 0.561 & 0.758 & 84.17 & 54.18 & 83.88 \\
MCIS~\cite{MCIS}& 0.756 & 0.783 & 84.02 & 43.58 & 83.85 & 0.557 & 0.747 & 84.97 & 53.85 & 84.34 \\
CGGM~\cite{CGGM}& 0.747 & 0.798 & 84.43 & 43.21 & 84.13 & 0.551 & 0.761 & 83.90 & 53.47 & 84.14 \\
DEVA~\cite{DEVA}& 0.730 & 0.787 & 84.40 & 46.33 & 84.45 & 0.541 & 0.769 & 83.17 & 52.28 & 83.75 \\
DLF~\cite{DLF}& 0.731 & 0.781 & 85.06 & 47.08 &85.04& 0.536 & 0.764 & 85.42 & 53.9 & 85.27 \\
EMOE~\cite{EMOE}& 0.710 & - & 85.4 & 47.7 & 85.4 & 0.536 & - & 85.3 & 54.1 & 85.3 \\
TSDA~\cite{TSDA}& 0.698&0.793&86.3&48.6&86.2 &0.534&0.767&86.15&54.67&86.09\\
\textbf{DBR (Ours)}&\textbf{0.681} &\textbf{0.811}& \textbf{86.87} & \textbf{49.26} & \textbf{86.83} & \textbf{0.526} & \textbf{0.788} & \textbf{86.73} &\textbf{55.62} & \textbf{86.78} \\

\bottomrule
\end{tabular}

\label{tab:main}

\end{table*}

\subsection{Bidirectional Rebalancing Fusion Module}

After TSF and AGPR, each modality has a shared representation and an updated private representation. By delaying strong fusion until after the two branches have been separately regularized, BRF can integrate complementary evidence without immediately reintroducing the same imbalance. To make this transition explicit, we derive a modality-level shared feature from the fused shared context and concatenate it with the updated private feature:
\begin{equation}
\mathbf{z}_m^{\mathrm{sha}} = \mathrm{Slice}_m(\mathbf{Z}^{\mathrm{fusion}}), \quad
\mathbf{F}_m = W_f [\mathbf{z}_m^{\mathrm{sha}}; \hat{\mathbf{z}}_m^{\mathrm{pri}}],
\end{equation}
where $\mathrm{Slice}_m(\cdot)$ denotes the segment associated with modality $m$ and $W_f$ is a learnable projection. BRF then models cross-modal dependencies and adaptively highlights informative features for prediction. We first concatenate all modality features to obtain a global embedding:
\begin{equation}
\mathbf{F}_{\mathrm{all}} = \mathrm{Concat}\big(\mathbf{F}_m\big)_{m \in \mathcal{M}}.
\end{equation}
BRF employs bidirectional cross-attention for each modality.

The forward path measures the influence of modality $m$ on the joint representation:
\begin{equation}
\mathbf{F}_{m \to \mathrm{all}} =
\mathrm{Softmax}\left(
\frac{\mathbf{F}_m W_Q (\mathbf{F}_{\mathrm{all}} W_K)^\top}{\sqrt{d_k}}
\right) (\mathbf{F}_{\mathrm{all}} W_V),
\end{equation}
and the backward path refines modality $m$ from the global context:
\begin{equation}
\mathbf{F}_{\mathrm{all} \to m} =
\mathrm{Softmax}\left(
\frac{\mathbf{F}_{\mathrm{all}} W'_Q (\mathbf{F}_m W'_K)^\top}{\sqrt{d_k}}
\right) (\mathbf{F}_m W'_V),
\end{equation}
where $W_Q, W_K, W_V, W'_Q, W'_K, W'_V$ are learnable projection matrices and $d_k$ is the key dimension. (Weights can be shared in practice; we keep the notation separate for clarity.)

The enhanced representation of modality $m$ is
\begin{equation}
\mathbf{Y}_m = \mathrm{LN}\big(
\mathbf{F}_m + \mathbf{F}_{m \to \mathrm{all}} + \mathbf{F}_{\mathrm{all} \to m}
\big).
\end{equation}
We further compute a global context embedding by pooling and averaging:
\begin{equation}
\bar{\mathbf{Y}} = \frac{1}{|\mathcal{M}|}
\sum_{m \in \mathcal{M}} \mathrm{Pool}(\mathbf{Y}_m),
\end{equation}
where $\mathrm{Pool}(\cdot)$ denotes temporal pooling (e.g., mean pooling).

Unlike methods that compute per-modality gates independently, we integrate the global context $\bar{\mathbf{Y}}$ into the gating calculation so that gates depend on both local and global cues. For each modality $m \in \mathcal{M}$, we compute
\begin{equation}
\psi_m =
\frac{\exp\left( \mathbf{q}^\top
\tanh\left( W_m [\mathbf{Y}_m ; \bar{\mathbf{Y}}] + \mathbf{bias}_m \right) \right)}
{\sum_{m' \in \mathcal{M}} \exp\left( \mathbf{q}^\top
\tanh\left( W_{m'} [\mathbf{Y}_{m'} ; \bar{\mathbf{Y}}] + \mathbf{bias}_{m'} \right) \right)},
\end{equation}
where $[\mathbf{Y}_m ; \bar{\mathbf{Y}}]$ is concatenation and $\mathbf{q}$, $W_m$, $\mathbf{bias}_m$ are learnable parameters. The final fused representation is
\begin{equation}
\mathbf{Y}_{\mathrm{fin}} = \sum_{m \in \mathcal{M}} \psi_m \odot \mathbf{Y}_m,
\end{equation}
where $\odot$ denotes element-wise multiplication.

\subsection{Objective Optimization}

For classification, we use cross-entropy loss; for regression, we adopt mean squared error as the main task loss $\mathcal{L}_{\text{task}}$. The overall objective integrates the task loss with regularization terms from the three auxiliary modules:
\begin{equation}
\mathcal{L}_{\text{all}} =
\mathcal{L}_{\text{task}}
+ \mathcal{L}_{\mathrm{MD}}
+ \mathcal{L}_{\mathrm{TSF}}
+ \mathcal{L}_{\mathrm{AGPR}}.
\end{equation}
By jointly optimizing these terms, DBR learns disentangled and complementary multimodal representations that reduce redundancy while preserving modality-specific cues for robust sentiment prediction.

\section{Experiments}

\subsection{Experimental Settings}

\textbf{Benchmarks and Metrics.} We evaluate DBR on three widely used multimodal benchmarks.
CMU-MOSI~\cite{Cmu-mosi} contains 2,199 monologue video segments with acoustic and visual features.
CMU-MOSEI~\cite{Cmu-mosei} consists of 22,856 video segments.
Both MOSI and MOSEI provide sentiment labels in the range $[-3, 3]$ (from strongly negative to strongly positive).
MIntRec~\cite{MIntRec} includes 2,224 samples across 20 intent categories. Following prior works~\cite{DEVA,EMOE}, we report Acc-2, Acc-7, F1-score, Mean Absolute Error (MAE), and Pearson Correlation (Corr) for MSA.
For Multimodal Intent Recognition (MIR) on MIntRec, we evaluate Accuracy, F1, Precision, and Recall.

\textbf{Implementation details.} All models are implemented in PyTorch and trained on a single NVIDIA A100 GPU (32GB).
The batch size is set to 128, weight decay to $1 \times 10^{-4}$, and the Adam optimizer is adopted.
We employ 5-fold cross-validation and early stopping with a patience of 6 epochs to ensure stable convergence.

\textbf{Model Zoo.} To verify the superiority of our proposed DBR, we experimentally compare it with the following state-of-the-art baseline models. Discourse vector fusion methods using tensor-based fusion and low-rank variants: TFN~\cite{TFN}, LMF~\cite{LMF}. Models that learn invariant and specific representations via feature decomposition: MISA~\cite{misa}, FDMER~\cite{FDMER}, ConFEDE~\cite{confede}, DMD~\cite{dmd},DLF~\cite{DLF}, EMOE~\cite{EMOE}, TSDA~\cite{TSDA}. Models that improve token representations using non-linguistic signals using attention mechanisms and Transformer modules: MuLT~\cite{MuLT}, PMR~\cite{PMR}, ALMT~\cite{ALMT}. Learning the multimodal and unimodal representations based on the multimodal label and generated unimodal labels: Self-MM~\cite{self-mm}, DEVA~\cite{DEVA}. Model which allows audio and video information to leak into the BERT model for multimodal fusion: MAG-BERT~\cite{MAG-BERT}.

\subsection{Comparison with State-of-the-art}

We select representative baselines spanning decomposition-based, interaction-based, and fusion-based families for a comprehensive comparison with DBR.

\textbf{Results on MSA benchmarks.}
Table~\ref{tab:main} presents the comparative results of DBR and several leading baselines on MOSI and MOSEI.
DBR achieves state-of-the-art performance across all evaluation metrics.
For example, on MOSI, DBR improves Acc-7 to 49.26\%, surpassing DLF by 2.18\%, and also outperforms recent methods such as EMOE and DEVA.
On MOSEI, DBR reaches an Acc-7 of 55.62\%, which is 1.60\% higher than the previous best baseline.
Consistent gains are also observed in Corr, Acc-2, MAE, and F1 on both datasets.

These improvements can be attributed to the collaborative effect of the four modules in DBR. MD provides a standard shared-private initialization. TSF explicitly separates temporal and structural evidence in the shared branch and suppresses redundancy caused by over-concentrated dominant patterns. AGPR preserves discriminative private cues and prevents private-feature dilution in the private branch. Finally, BRF integrates the two regularized branches and highlights informative features for final prediction.

\textbf{Results on the MIR benchmark.}
Table~\ref{tab:MIR} reports MIR performance on MIntRec.
DBR outperforms strong baselines on all four metrics.
In particular, DBR achieves 73.04\% accuracy and 72.47\% F1-score, clearly surpassing EMOE and other recent methods, which confirms that the proposed framework generalizes well beyond sentiment regression/classification to multimodal intent recognition.

\begin{table}[t]
  \centering
  \caption{Performance comparison on MIntRec (\%).}
\setlength{\tabcolsep}{6pt}
\renewcommand{\arraystretch}{0.9}
  \begin{tabular}{lcccc}
    \toprule
    \textbf{Models} & \textbf{Acc.} & \textbf{F1} & \textbf{Pre.} & \textbf{Rec.} \\
    \midrule
    \midrule
    MAG-BERT~\cite{MAG-BERT}& 70.34 & 68.19 & 68.31 & 69.36 \\
    MuLT~\cite{MuLT}        & 72.58 & 69.36 & 70.73 & 69.47 \\
    MISA~\cite{misa}        & 72.36 & 70.57 & 71.24 & 70.41 \\
    GsiT~\cite{GsiT}        & 72.60 & 69.40 & 69.40 & 70.10 \\
    CAGC~\cite{CAGC}        & 72.53 & 70.62 & 70.86 & 70.55 \\
    EMOE~\cite{EMOE}        & 72.58 & 70.73 & 72.08 & 70.86 \\
    TSDA~\cite{TSDA}        & 72.67 & 70.79 & 72.19 & 70.85 \\
    \textbf{DBR (Ours)}    & \textbf{73.04} & \textbf{72.47} & \textbf{73.03} & \textbf{72.37} \\
    \bottomrule
  \end{tabular}
   \label{tab:MIR}
\end{table}

\begin{table}[t]
\caption{Ablation studies of DBR on MOSI and MOSEI.}
\centering
\setlength{\tabcolsep}{5pt}
\renewcommand{\arraystretch}{0.9}
\begin{tabular}{l|cc|cc}
\toprule
\multirow{2}{*}{Model}
    & \multicolumn{2}{c|}{CMU-MOSI}
    & \multicolumn{2}{c}{CMU-MOSEI} \\
    & MAE $\downarrow$ & Corr $\uparrow$ & MAE $\downarrow$ & Corr $\uparrow$ \\
\midrule
\midrule
\textbf{DBR (Ours)} & \textbf{0.681} & \textbf{0.811} & \textbf{0.526} & \textbf{0.788} \\
\midrule
\multicolumn{5}{c}{\textit{\textbf{Importance of DBR Components}}} \\
w/o TSF (Ours)    & 0.726 & 0.784 & 0.560 & 0.759  \\
w/o AGPR (Ours)  & 0.732 & 0.773 & 0.577 & 0.746  \\
w/o BRF (Ours)   & 0.724 & 0.779 & 0.544 & 0.749 \\
\midrule
\multicolumn{5}{c}{\textit{\textbf{Importance of Modality}}} \\
w/o Linguistic   & 1.036 & 0.382 & 0.845 & 0.396  \\
w/o Acoustic     & 0.873 & 0.731 & 0.589 & 0.732 \\
w/o Visual       & 0.918 & 0.709 & 0.620 & 0.725 \\
\midrule
\multicolumn{5}{c}{\textit{\textbf{Different Fusion Mechanisms}}} \\
Addition         & 0.746 & 0.762 & 0.563 & 0.747  \\
Multiplication   & 0.741 & 0.771 & 0.542 & 0.753 \\
CMAF~\cite{FDMER}& 0.729 & 0.785 & 0.533 & 0.766 \\
\midrule
\multicolumn{5}{c}{\textit{\textbf{Importance of Regularization}}} \\
w/o $\mathcal{L}_{\mathrm{MD}}$ & 0.726 & 0.785 & 0.562 & 0.756 \\
w/o $\mathcal{L}_{\mathrm{TSF}}$ & 0.751 & 0.763 & 0.570 & 0.743 \\
w/o $\mathcal{L}_{\mathrm{AGPR}}$ & 0.760 & 0.758 & 0.577 & 0.732 \\
CEntropy Loss                   & 0.752 & 0.763 & 0.572 & 0.744 \\
\bottomrule
\end{tabular}
\label{tab:ablation-2}
\end{table}

\subsection{Ablation Studies}

To thoroughly examine the effectiveness of DBR, we conduct ablation experiments as summarized in Table~\ref{tab:ablation-2}.

\textbf{Importance of DBR components.}
We first remove each major component in turn.
Without the TSF module, MAE increases by about 0.03 and Corr drops by around 0.02 on both datasets, showing that explicit modeling of temporal-structural structure is important for robust sentiment prediction.
Removing the AGPR (Anchor-Guided Private) module leads to the largest performance degradation (MAE +0.04, Corr $-0.03$), highlighting the necessity of preserving modality-specific cues and preventing private-space homogenization.
Removing the BRF (Bidirectional Rebalancing Fusion) module also degrades performance (MAE +0.02, Corr $-0.03$), indicating its role in optimizing the integration of shared and private representations.
Overall, TSF, AGPR, and BRF are complementary and jointly address information redundancy and private feature homogeneity.

\textbf{Importance of modality.}
We further analyze the contribution of each modality by removing one modality at a time and training bimodal DBR variants.
Excluding the linguistic modality results in a dramatic performance drop, confirming that text carries the most reliable sentiment signal.
Removing acoustic or visual features also causes notable degradation, although less severe than removing text.
In all cases, the trimodal configuration consistently outperforms its bimodal counterparts, indicating that each modality provides indispensable complementary information.

\textbf{Different fusion mechanisms.}
We compare BRF with several alternative fusion strategies. BRF outperforms additive fusion, multiplicative fusion, and CMAF~\cite{FDMER} on both MOSI and MOSEI, demonstrating the benefit of explicitly modeling bidirectional cross-modal interactions and context-aware gating rather than relying on fixed or shallow fusion operators.

\textbf{Importance of regularization.}
To assess the role of each regularizer, we ablate them individually.
Removing $\mathcal{L}_{\mathrm{MD}}$ causes a small Corr drop, indicating that redundancy increases but the model can still partially rely on other modules.
Removing $\mathcal{L}_{\mathrm{TSF}}$ leads to a larger decline, as temporal and structural cues become entangled and mutual interference is amplified.
Removing $\mathcal{L}_{\mathrm{AGPR}}$ yields the largest degradation, showing that suppressing private-space homogenization and preserving modality-specific uniqueness is most critical for performance.
Replacing these targeted regularizers with a global contrastive CEntropy loss leads to moderate performance decay: while contrastive supervision helps maintain class separation, it cannot fully prevent redundancy and private-modality homogenization.


\subsection{Qualitative Analysis}
\begin{figure}[htbp]
    \centering
    \includegraphics[width=1\linewidth]{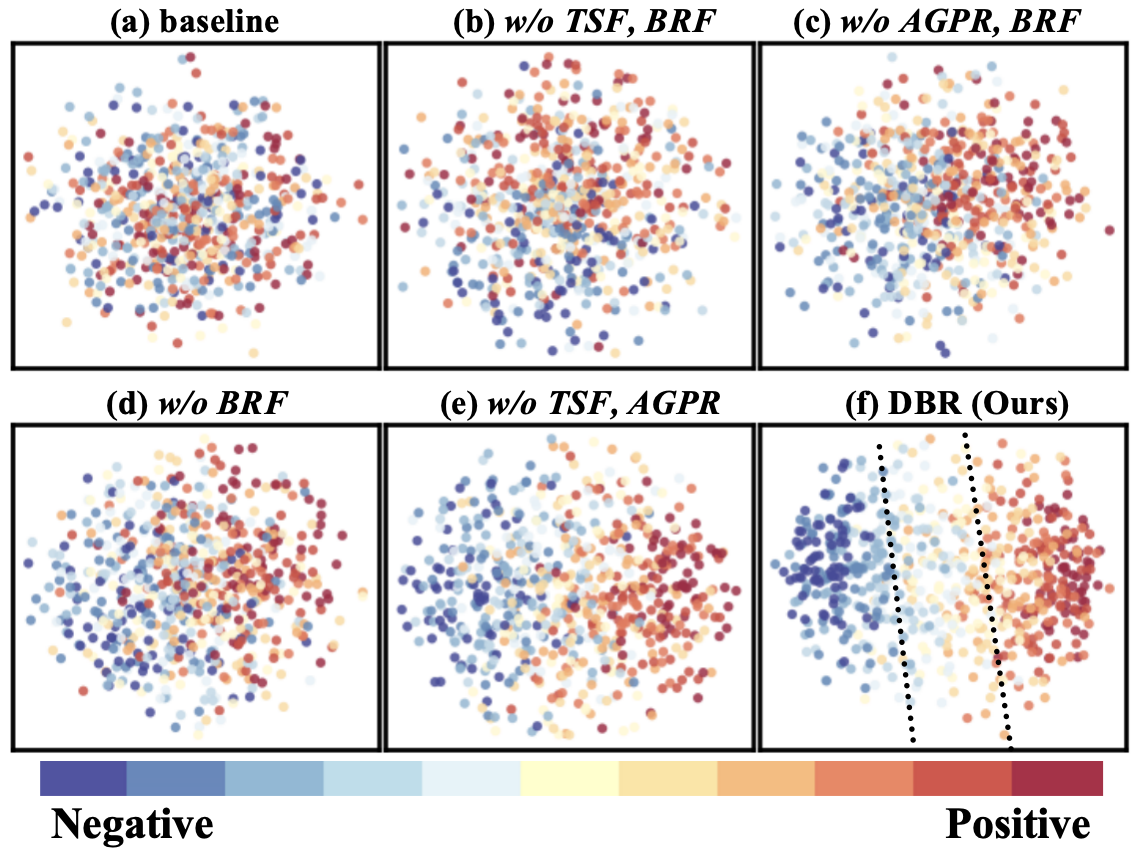}
    \caption{t-SNE visualization of feature distributions on MOSI. Warmer colors indicate more positive sentiment. The full DBR variant produces a clearer sentiment gradient.}
    \Description{A t-SNE plot comparing multiple ablation variants on MOSI. The full DBR model forms a smoother sentiment gradient and more coherent structure than variants missing TSF, AGPR, or BRF.}
    \label{fig:TSNE}
\end{figure}

We qualitatively evaluate DBR with t-SNE visualizations of the MOSI test features.
As shown in Fig.~\ref{fig:TSNE}, the variant lacking TSF, AGPR, and BRF produces irregular and dispersed feature distributions, indicating weak structure and poor sentiment separation.
Incorporating TSF or AGPR individually leads to noticeable improvements, but overlaps between different sentiment levels still remain.
When BRF is removed, the structure becomes clearer, yet some sentiment scores are still mixed.
The complete DBR model yields a compact, continuous, and well-ordered gradient in the feature space.
Since MSA is formulated as a regression task, the ideal outcome is a smooth sentiment continuum rather than discrete clusters, which is consistent with our visualization.

\subsection{Analysis of attention distributions}
\begin{figure}[htbp]
    \centering
    \includegraphics[width=0.9\linewidth]{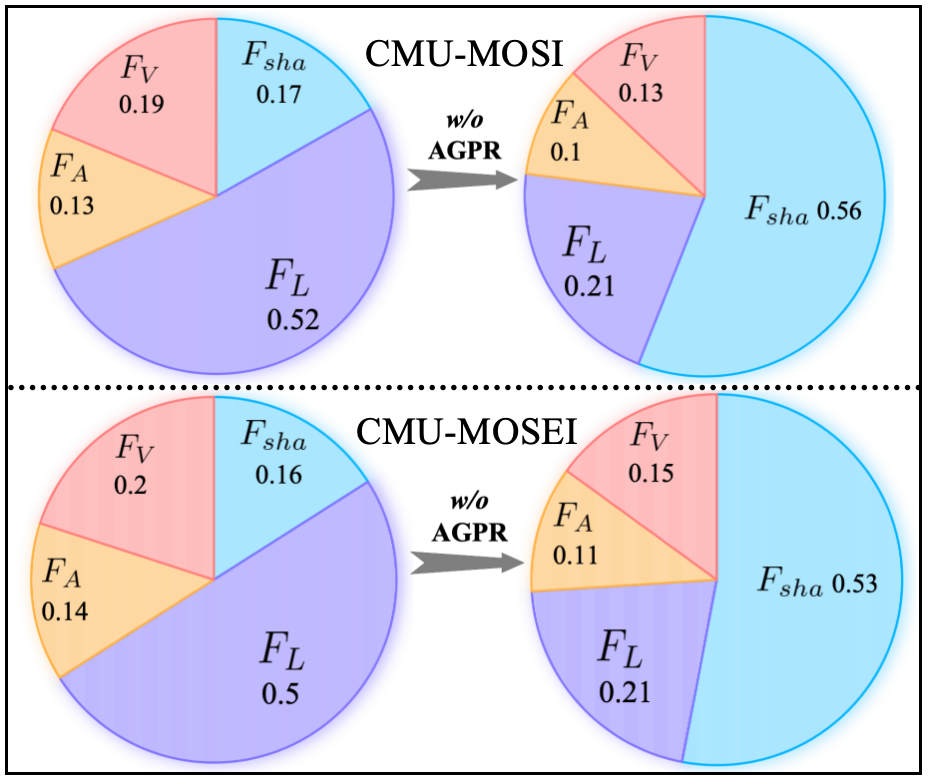}
    \caption{Attention distributions from the BRF module on the MOSI (top) and CMU-MOSEI benchmark (bottom) test set. Numbers denote attention values. $F_{sha}$, $F_A$, $F_L$, $F_V$ are shared, acoustic, linguistic and visual components, respectively.}
    \Description{A bar-chart style visualization of attention scores assigned to shared and private features across modalities on MOSEI, showing that private features, especially linguistic ones, receive larger weights.}
    \label{fig:atten-all}
\end{figure}

Fig.~\ref{fig:atten-all} presents the attention distributions from the BRF module on CMU-MOSI (top) and CMU-MOSEI (bottom). Across both datasets, attention scores assigned to modality-specific (private) representations are consistently higher than those of modality-invariant (shared) representations, especially for the linguistic modality. This suggests that private features play a more critical role in capturing fine-grained emotional cues. When the AGPR module is removed, attention on private representations clearly decreases and falls below that of shared representations on both datasets. In this case, private features lose discriminative uniqueness, and the model is forced to rely mainly on shared information. These consistent patterns across both datasets further highlight the importance of maintaining distinct modality-specific representations and validate the role of AGPR in preventing modality homogenization.

\begin{figure}[htbp]
    \centering
    \includegraphics[width=1\linewidth]{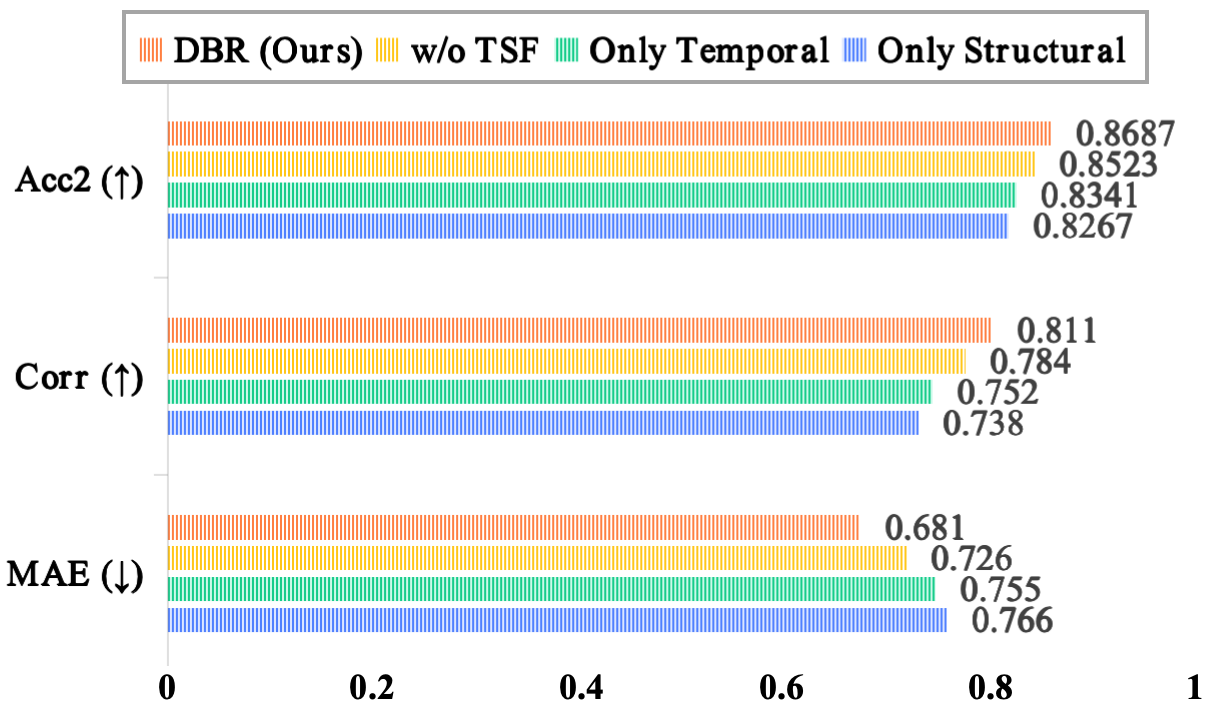}
    \caption{Analysis of temporal-structural factorization on MOSI.}
    \Description{A performance comparison across full DBR, a version without TSF, and temporal-only or structural-only variants, showing that explicit temporal-structural factorization performs best.}
    \label{fig:STDecouple}
\end{figure}

\subsection{Analysis of temporal-structural factorization}
To verify the effectiveness of the temporal-structural factorization, we conduct additional ablations on MOSI.
As shown in Fig.~\ref{fig:STDecouple}, the full DBR model with TSF achieves the best performance on Acc2, Corr, and
MAE. Removing TSF (w/o TSF) still performs better than using temporal-only or structural-only features.
A plausible explanation is the inherent complementarity between temporal dynamics (e.g., pitch changes, facial motion) and structural attributes (e.g., sentiment-bearing words, static visual cues).
Even without explicit decoupling, self-attention can capture part of both dimensions.
However, relying on only temporal or only structural representations causes severe information loss and substantially weakens the mutual information between representations and labels.

\subsection{Visualization of module weights and importance.}
\begin{figure}[t]
    \centering
    \includegraphics[width=\linewidth]{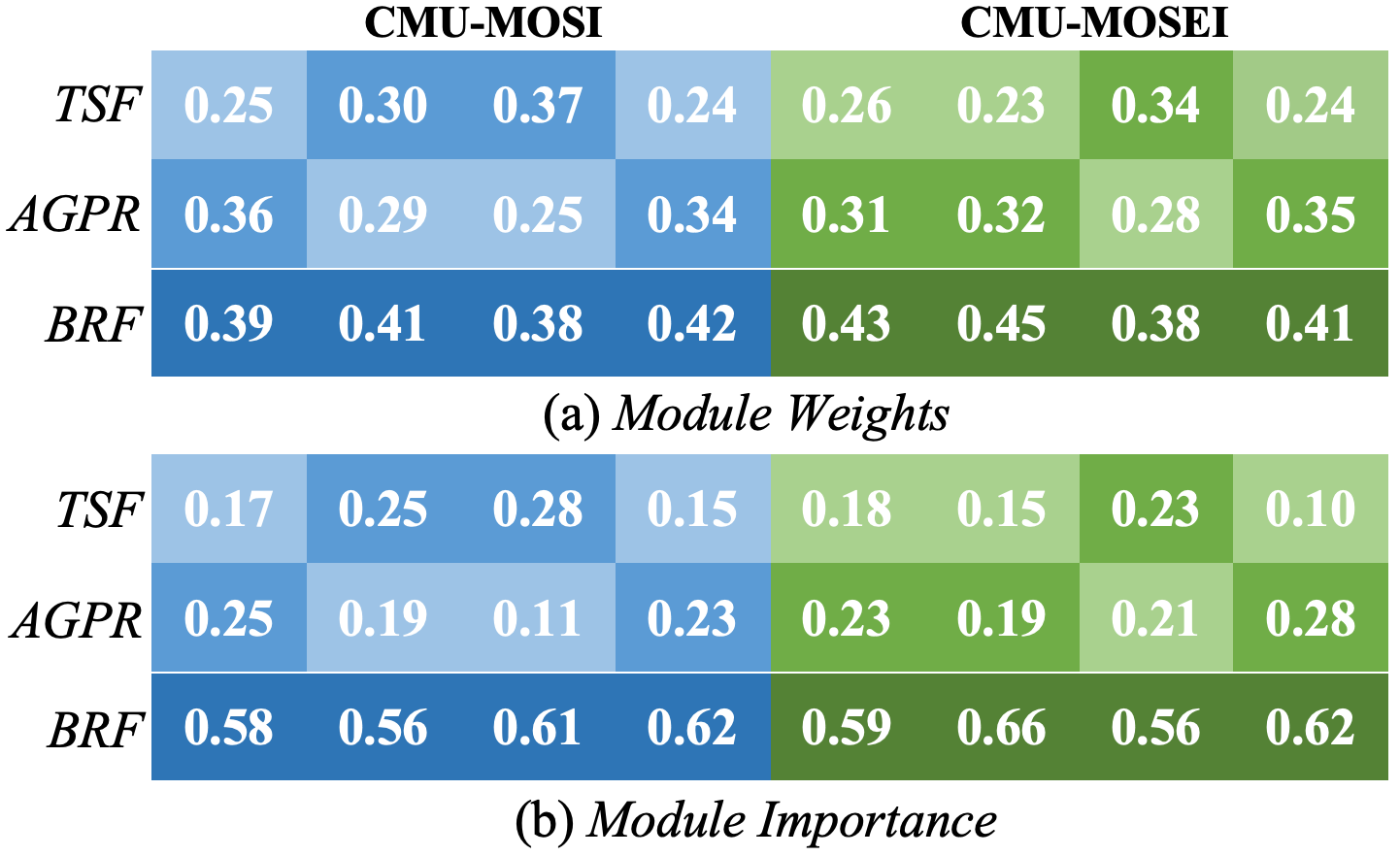}
    \caption{Average fusion weights and estimated contributions of the three cooperative modules in DBR on MOSI and MOSEI. Higher values indicate more important modules.}
    \Description{Two grouped bar charts summarizing the learned weights and estimated contributions of the MD, TSF, AGPR, and BRF spaces on MOSI and MOSEI. BRF and the branch-specific modules contribute more than MD.}
    \label{fig:weight}
\end{figure}

To further understand how DBR exploits its three cooperative modules, we visualize the learned fusion weights and the estimated contributions of each module on MOSI and MOSEI, as shown in Fig.~\ref{fig:weight}.
BRF receives the highest average weight and contribution on both datasets, indicating that final decisions are mainly driven by the module that integrates shared and private representations.
TSF and AGPR also obtain consistently high weights, confirming that explicit modeling of temporal-structural organization and modality-specific cues provides strong complementary evidence.
These patterns are consistent with the ablation results in Table~\ref{tab:ablation-2}, and the agreement between learned weights and measured contributions supports both the effectiveness and interpretability of the multi-module collaborative design in DBR.

\subsection{Sensitivity Analysis}
\begin{figure}[htbp]
    \centering
    \includegraphics[width=1\linewidth]{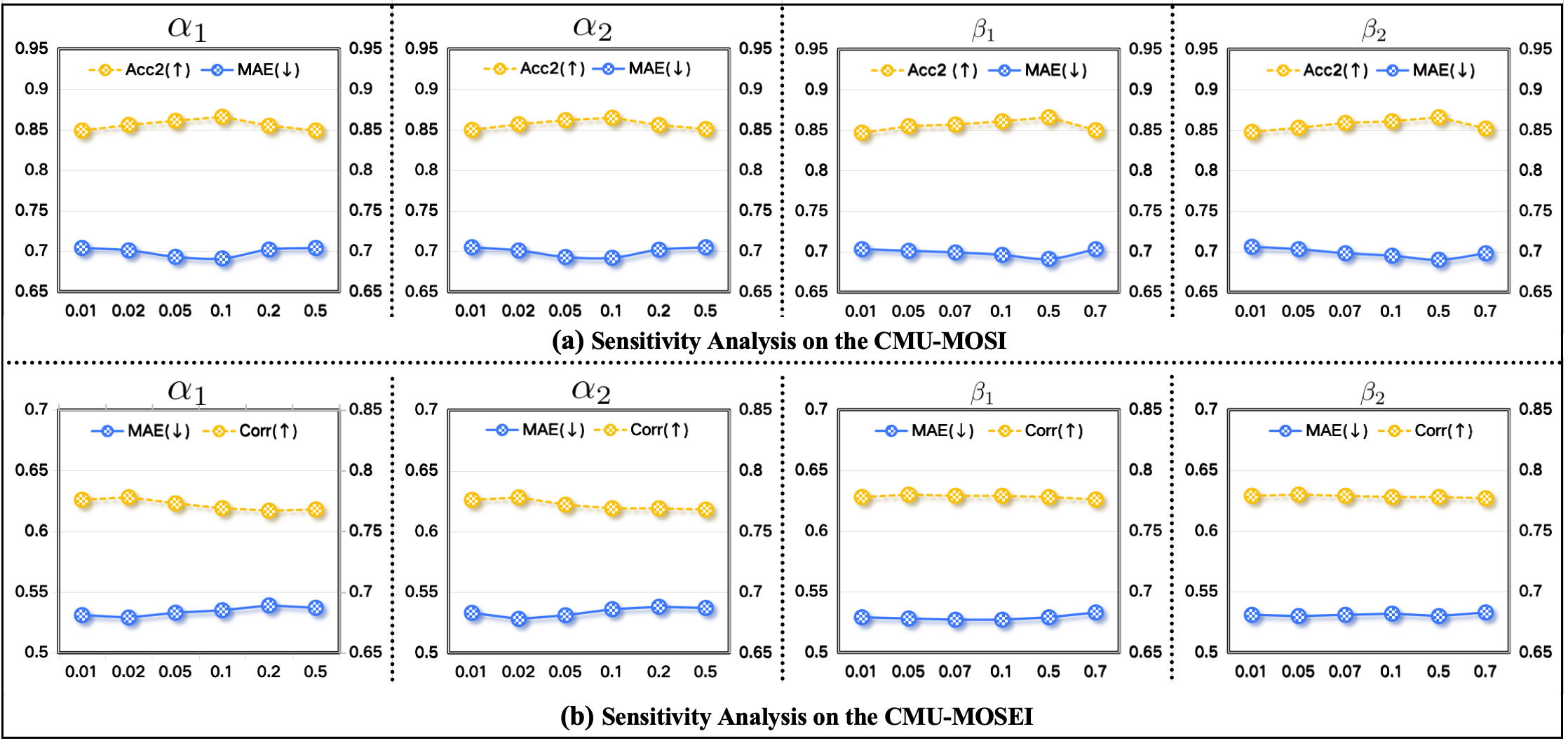}
    \caption{Sensitivity analysis of DBR on MOSI and MOSEI.}
    \Description{Line charts showing that evaluation metrics remain stable as the regularization hyperparameters and routing temperature vary within a reasonable range on MOSI and MOSEI.}
    \label{fig:sen}
\end{figure}

We finally assess the robustness of DBR by performing sensitivity analysis on the main regularization hyperparameters $\alpha_1$, $\alpha_2$, $\beta_1$, and $\beta_2$, varying each within a reasonable range while keeping the others fixed on MOSI and MOSEI.
As shown in Fig.~\ref{fig:sen}, both classification and regression metrics remain highly stable: on MOSEI, MAE stays between 0.526 and 0.531, and Corr ranges from 0.776 to 0.780, with similar trends on MOSI.
We also examine the routing temperature $\gamma$ and only observe instability under extreme values.
These results indicate that DBR is not sensitive to precise hyperparameter choices and maintains consistent performance without extensive tuning, which is beneficial for practical deployment and reproducible multimodal learning.

\begin{table}[h]
\setlength{\tabcolsep}{1.0pt}
\renewcommand{\arraystretch}{1}
\centering
\caption{Efficiency Analysis on CMU-MOSEI.}
\label{tab:efficiency_metrics}
\begin{tabular}{ccccc}
\hline
Model& Params(M)& Time/Epoch(s) &F1(\%)&MAE ($\downarrow$)\\ \hline
MISA~\cite{misa}& 112.45  & 14.31 & 82.13&0.594 \\
DMD~\cite{dmd}     & 122.06 & 35.97  & 83.88&0.561 \\
EMOE~\cite{EMOE}& 128.60  & 21.86 & 85.3&0.536 \\ 
DBR (Ours) & 127.18  & 20.76 & \textbf{86.78}&\textbf{0.526} \\ \hline
\end{tabular}
\label{tab:Efficiency Analysis}
\end{table}

\subsection{Efficiency Analysis}
DBR achieves a superior trade-off between predictive performance and computational overhead, as shown in Table~\ref{tab:Efficiency Analysis}. Compared
with heavier decomposition-based baselines such as EMOE~\cite{EMOE} and DMD~\cite{dmd}, DBR achieves stronger predictive performance across F1 and MAE with lower per-epoch time. Although TSF and BRF introduce additional computation, the main operations are performed on already encoded modality features rather than on raw
inputs, so the practical overhead remains moderate. This result is
important for the overall positioning of DBR: the gain does not
simply come from scaling model size, but from a more structured
organization of shared and private evidence.

\section{Conclusion}

This paper proposes DBR to mitigate heterogeneity-induced shared-private branch imbalance in multimodal sentiment analysis. Built on a standard shared-private initialization, DBR shows that the persistent difficulty can be understood through two coupled manifestations: redundant shared evidence and private-feature dilution. The Temporal-Structural Factorization module regulates the shared branch, the Anchor-Guided Private Routing module preserves discriminative private cues while allowing controlled borrowing, and the Bidirectional
Rebalancing Fusion module reunifies the two branches for prediction. Experiments on multiple benchmarks confirm that this coordinated design yields strong performance and generalization.







\bibliographystyle{ACM-Reference-Format}
\bibliography{software_revised}

\end{document}